\documentclass[prd,aps,preprintnumbers,amsmath,amssymb,nofootinbib]{revtex4-2}

\usepackage{graphicx}
\usepackage{bm}
\usepackage{physics}
\usepackage{mathtools}
\usepackage[breaklinks,colorlinks=true,linkcolor=blue,citecolor=blue,urlcolor=blue]{hyperref}

\begin{document}

\title{Energy-momentum tensor at small $x$ \\
from gravitational transverse-momentum distributions}

\author{Tomoya Uji}
\email{uji@nt.phys.s.u-tokyo.ac.jp}
\affiliation{Department of Physics, The University of Tokyo, 7-3-1 Hongo, Bunkyo-ku, Tokyo 113-0033, Japan}

\begin{abstract}
We investigate the momentum-space mechanical properties of small-$x$ gluons and sea quarks using gravitational transverse-momentum distributions (gravitational TMDs).  For gluons, we formulate these distributions using the light-front gauge-invariant canonical energy-momentum tensor and derive their relations to ordinary TMD correlators, identifying momentum densities, the transverse flux of inertia, and transverse stresses.  In the eikonal approximation, the longitudinal and transverse canonical gluon momentum densities are determined by the unpolarized distribution, whereas the transverse flux of inertia and transverse stresses vanish.  For an unpolarized target, we numerically evaluate the gluon and sea-quark momentum densities using a common initial condition based on the McLerran--Venugopalan model and fixed-coupling leading-order Balitsky--Kovchegov evolution.  We also evaluate the sea-quark transverse flux of inertia through the twist-three quark TMD at the high transverse-momentum region.  These results illustrate departures from the free-particle relation between canonical transverse momentum and the effective transverse velocity governing inertia flow.
\end{abstract}

\maketitle

\section{Introduction}

Understanding the internal properties of hadrons in terms of quarks and gluons is a central goal of Quantum Chromodynamics (QCD).  The Electron--Ion Collider (EIC) will extend the experimental investigation of the multidimensional structure of hadrons, providing new opportunities to study the distributions and correlations of partons~\cite{Accardi:2012qut, AbdulKhalek:2021gbh}.  An important aspect of this program concerns the mechanical structure of hadrons: how energy and momentum are distributed among their constituents, and how internal forces arise from their interactions~\cite{Polyakov:2018zvc, Burkert:2023wzr}.  These questions are naturally formulated in terms of the QCD energy-momentum tensor (EMT), which describes energy and momentum densities together with their fluxes and stresses~\cite{Lorce:2018egm}.

The off-forward matrix elements of the local EMT are parametrized by gravitational form factors (GFFs).  Their connection to moments of generalized parton distributions (GPDs) provides access to information about the EMT through hard exclusive processes~\cite{Ji:1996ek, Polyakov:2002yz}.  In particular, the $D$ form factor encodes information about internal stress and, through appropriate spatial representations, the distributions of pressure and shear forces~\cite{Polyakov:2002yz, Polyakov:2018zvc}.  Experimental analyses of deeply virtual Compton scattering and near-threshold quarkonium production have further stimulated the study of nucleon mechanical properties and gluonic GFFs~\cite{Burkert:2018bqq, Duran:2022xag}.  These studies primarily address the spatial organization of the hadron through its dependence on momentum transfer.  A complementary question is how its mechanical properties are distributed over the longitudinal momentum fraction $x$ and intrinsic transverse momentum $\bm{k}_T$ of its constituents.  This momentum resolution concerns a different variable from the momentum transfer underlying spatial imaging.

A broader description can be obtained by retaining both position and momentum information.  Generalized transverse-momentum-dependent distributions (GTMDs) parametrize off-forward light-front parton correlators with intrinsic transverse momentum retained~\cite{Meissner:2009ww, Lorce:2013pza}.  At zero skewness, Fourier transforms of the corresponding correlators with respect to transverse momentum transfer define partonic Wigner distributions~\cite{Lorce:2011kd}.  Husimi distributions, obtained by Gaussian smearing of Wigner distributions over transverse position and momentum, have also been proposed as a description of partonic phase space at finite resolution~\cite{Hagiwara:2014iya, Hagiwara:2016kam}.  This phase-space perspective also suggests a connection with kinetic theory~\cite{Belitsky:2003nz}, where the EMT of freely propagating on-shell particles is obtained by integrating a phase-space distribution weighted by $k^\mu k^\nu$ with the appropriate momentum measure~\cite{Wu:2017rry, Weickgenannt:2019dks}.  As a first step toward a phase-space description of these mechanical properties in the light front, we focus on the EMT resolved in the parton longitudinal momentum fraction and transverse momentum, with the transverse position integrated out.

The EMT in momentum space can be parametrized in terms of gravitational transverse-momentum distributions (gravitational TMDs), whose relations to ordinary quark TMDs have been established~\cite{Lorce:2023zzg}.  These relations provide a mechanical interpretation of the relevant TMD projections in terms of momentum densities, the transverse flux of inertia, and stress distributions.  Quark gravitational TMDs and the associated momentum-space mechanical properties have been investigated in phenomenological quark--diquark models~\cite{Saha:2026fwe, Kaur:2026wxm, Kaur:2026jjr}.  A gluon construction based on the kinetic EMT has also been developed~\cite{Saha:2026veh}.  In this work, we adopt the light-front gauge-invariant canonical (gic) gluon EMT~\cite{Lorce:2015lna} for a direct interpretation in terms of canonical partonic momentum, following the quark construction of Ref.~\cite{Lorce:2023zzg}.

In this work, we apply gravitational TMDs to small-$x$ gluons and sea quarks to investigate their mechanical properties.  Studies of helicity, orbital angular momentum, and spin--orbit correlations have revealed nontrivial spin and orbital structure in this regime~\cite{Hatta:2016aoc, Kovchegov:2025gcg, Bhattacharya:2024sno}.  The energy-momentum structure has also been investigated through the gluon GFF $A(t)$~\cite{Hagiwara:2024wqz}.  Gravitational TMDs complement these studies by resolving the EMT in the parton longitudinal momentum fraction and transverse momentum, providing a description of small-$x$ systems in terms of mechanical properties.

We use the metric $g^{\mu\nu} = \operatorname{diag}(1, -1, -1, -1)$ and two lightlike vectors satisfying $n^2 = \bar{n}^2 = 0$ and $n\cdot\bar{n} = 1$.  We decompose a four-vector as $v^\mu = v^+\bar{n}^\mu + v^-n^\mu + v_T^\mu$, where $v^+ = v\cdot n$ and $v^- = v\cdot\bar n$.  We use the light-front conventions
\begin{equation}
    v^\pm = \frac{v^0 \pm v^3}{\sqrt{2}}\,, \qquad g^{+-} = 1\,, \qquad g^{ij} = - \delta^{ij}\,,
\end{equation}
and choose
\begin{equation}
    P^\mu = \left[P^+, \frac{M^2}{2P^+}, \bm{0}_T\right]\,, \qquad k^\mu = \left[xP^+, k^-, \bm{k}_T\right]\,, \qquad n^\mu = \left[0, 1, \bm{0}_T\right]\,.
\end{equation}
Throughout this paper, $P^\mu$ and $k^\mu$ denote the four-momenta of the parent nucleon and the parton, respectively.

\section{Gauge-invariant canonical gravitational TMDs of gluons}
\label{sec:gluon-gTMDs}

We formulate gluon gravitational TMDs using the light-front gic EMT.  After introducing their operator definition and Lorentz parametrization, we derive their relations to ordinary gluon TMD correlators.

\subsection{Gauge-invariant canonical EMT of gluons}
\label{subsec:gluon-gic-EMT}

Gluon gravitational TMDs have recently been studied using the kinetic (Belinfante--Rosenfeld) EMT~\cite{Saha:2026veh}.  In the present work, however, we adopt the light-front gic EMT for a direct interpretation in terms of partonic momentum densities, following the quark construction of Ref.~\cite{Lorce:2023zzg}.  The gluon gic EMT is given by~\cite{Lorce:2015lna}
\begin{equation}
    T_{g, \mathrm{gic}}^{\mu\nu}(r; \eta) = - \frac{1}{2}\left\{F_a^{\mu\alpha}(r), \left(D_{\mathrm{pure}}^\nu A_{\mathrm{phys}, \alpha}\right)_a(r)\right\} + \frac{1}{4}g^{\mu\nu}F_a^{\alpha\beta}(r)F_{a, \alpha\beta}(r)\,,
\label{eq:gluon-gic-EMT}
\end{equation}
where $F_a^{\mu\nu}$ is the gluon field-strength tensor and $\{X, Y\} = XY + YX$ denotes the operator anticommutator, introduced to ensure hermiticity.  

The gauge field is separated into pure-gauge and physical components as~\cite{Chen:2008ag}
\begin{equation}
    A^\mu = A_{\mathrm{pure}}^\mu + A_{\mathrm{phys}}^\mu\,.
    \label{eq:gauge-field-separation}
\end{equation}
Here, $A_{\mathrm{pure}}^\mu$ transforms as a gauge connection, while $A_{\mathrm{phys}}^\mu$ transforms covariantly.  These transformation properties ensure the gauge invariance of Eq.~\eqref{eq:gluon-gic-EMT}.  The covariant derivative constructed from the pure-gauge field $A_{\mathrm{pure}}^\mu$ acts on a field $X = X_at^a$ as
\begin{equation}
    D_{\mathrm{pure}}^\mu X = \partial^\mu X - ig[A_{\mathrm{pure}}^\mu, X]\,.
    \label{eq:adjoint-pure-derivative}
\end{equation}
To interpret the partonic momentum as canonical momentum, we further impose $F_{\mathrm{pure}}^{\mu\nu} = 0$, where $F_{\mathrm{pure}}^{\mu\nu}$ is the field strength constructed from $A_{\mathrm{pure}}^\mu$.  This condition ensures $[D_{\mathrm{pure}}^\mu, D_{\mathrm{pure}}^\nu] = 0$, allowing $iD_{\mathrm{pure}}^\mu$ to serve as a gauge-covariant canonical momentum operator~\cite{Lorce:2012ce}.

The parameter $\eta = +1$ ($-1$) denotes a future-pointing (past-pointing) gauge link and fixes the associated boundary prescription~\cite{Hatta:2011ku, Lorce:2012ce}.  For the bilocal gluon correlators introduced below, we adopt the Weizs\"acker--Williams (WW) gauge-link structure, with both fundamental staple links passing through the same light-front infinity, motivated by its partonic interpretation as in Ref.~\cite{Hatta:2016aoc}.  For this light-front construction, $A_{\mathrm{phys}}^+ = 0$.  In the gauge $A^+ = 0$ with the matching boundary conditions, $A_{\mathrm{pure}}^\mu = 0$ and $A_{\mathrm{phys}}^\mu = A^\mu$, so that Eq.~\eqref{eq:gluon-gic-EMT} reduces to the canonical gluon EMT~\cite{Hatta:2011zs}.  In a general gauge, the operator remains gauge invariant while retaining the nonlocality associated with the Wilson-line prescription~\cite{Lorce:2015lna}.

\subsection {Transverse-momentum-dependent gic EMT and gravitational TMDs of gluons}
To resolve the gluon momentum, we introduce a gauge-invariant bilocal extension of Eq.~\eqref{eq:gluon-gic-EMT}, with the field operators evaluated at $r_\pm = r \pm z / 2$, 
\begin{equation}
\begin{split}
    \widetilde{T}_{g, \mathrm{gic}}^{\mu\nu}(r, z; \eta) &= - \frac{1}{2}\left[F_a^{\mu\alpha}(r_-)\mathcal{U}_\eta^{ab}\left(D_{\mathrm{pure}}^\nu A_{\mathrm{phys}, \alpha}\right)_b(r_+) + \left(D_{\mathrm{pure}}^\nu A_{\mathrm{phys}, \alpha}\right)_a(r_-)\mathcal{U}_\eta^{ab}F_b^{\mu\alpha}(r_+)\right] \\
    &\qquad + \frac{1}{4}g^{\mu\nu}F_a^{\alpha\beta}(r_-)\mathcal{U}_\eta^{ab}F_{b, \alpha\beta}(r_+)\,.
\end{split}
\label{eq:gluon-gic-bilocal}
\end{equation}
Here, $\mathcal{U}_\eta^{ab}\coloneqq\mathcal{U}_\eta^{ab}(r_-, r_+)$ denotes a staple-shaped Wilson line in the adjoint representation, connecting $r_-$ and $r_+$ via light-front infinity in the direction $\eta n^\mu$, with the connection at infinity understood.

Fourier transformation with respect to the relative separation, $z$, defines the phase-space EMT operator,
\begin{equation}
    T_{g, \mathrm{gic}}^{\mu\nu}(r, k; \eta) = \int\frac{d^4z}{(2\pi)^4}\,e^{ik\cdot z}\widetilde{T}_{g, \mathrm{gic}}^{\mu\nu}(r, z; \eta)\,,
    \label{eq:gluon-gic-phase-space}
\end{equation}
whose integral over $k$ recovers the coincident-point EMT in Eq.~\eqref{eq:gluon-gic-EMT}.

For a hadron with momentum $P$ and spin $S$, the forward matrix element $\Theta_g^{\mu\nu}(P, k, S; \eta) = \mel{P, S}{T_{g, \mathrm{gic}}^{\mu\nu}(0, k; \eta)}{P, S}$ defines the fully unintegrated EMT correlator.  Integrating over the gluon light-front energy $k^-$ yields the transverse-momentum-dependent EMT,
\begin{equation}
\begin{split}
    \mathcal{T}_g^{\mu\nu}(P, x, \bm{k}_T, S; \eta) &= \int dk^-\,\Theta_g^{\mu\nu}(P, k, S; \eta) \\
    &= \int\frac{dz^-d^2\bm z_T}{(2\pi)^3}\,e^{ixP^+z^- - i\bm{k}_T\cdot\bm{z}_T}\mel{P, S}{\widetilde{T}_{g, \mathrm{gic}}^{\mu\nu}(0, z; \eta)}{P, S}\eval_{z^+ = 0}\,.
\end{split}
\label{eq:gluon-gravitational-TMD}
\end{equation}
The $k^-$ integration restricts the field insertions to equal light-front time, $z^+ = 0$; all endpoint derivatives in Eq.~\eqref{eq:gluon-gic-bilocal} are evaluated before imposing this restriction.  Integration over $x$ and $\bm{k}_T$ gives
\begin{equation}
    \int_{-\infty}^{\infty}dx\int d^2\bm{k}_T\,\mathcal{T}_g^{\mu\nu}(P, x, \bm{k}_T, S; \eta) = \frac{1}{P^+}\mel{P, S}{T_{g, \mathrm{gic}}^{\mu\nu}(0; \eta)}{P, S}\,.
\label{eq:gluon-gravitational-TMD-moment}
\end{equation}
This relation connects the momentum-resolved distribution to the forward matrix element of the EMT.

Following the general treatment of the light-front gauge-invariant EMT~\cite{Lorce:2015lna, Lorce:2023zzg}, we parametrize $\mathcal{T}_g^{\mu\nu}$ in terms of the available four-vectors and tensors, subject to parity, hermiticity, and time-reversal symmetry.  The resulting Lorentz decomposition reads~\cite{Lorce:2023zzg}
\begin{equation}
\begin{aligned}
    \mathcal{T}_g^{\mu\nu}(P, x, \bm{k}_T, N, S; \eta) &= \frac{1}{P^+}\biggl\{P^\mu P^\nu a_1^g + N^\mu N^\nu a_2^g + k_T^\mu k_T^\nu a_3^g + P^\mu N^\nu a_4^g + N^\mu P^\nu a_5^g \\
    &\qquad\qquad\qquad\qquad + P^\mu k_T^\nu a_6^g + k_T^\mu P^\nu a_7^g + N^\mu k_T^\nu a_8^g + k_T^\mu N^\nu a_9^g + M^2g_T^{\mu\nu}a_0^g
    \\
    &\qquad - \frac{\epsilon_T^{k_TS_T}}{M}\Bigl[P^\mu P^\nu a_{1T}^{\perp g} + N^\mu N^\nu a_{2T}^{\perp g} + k_T^\mu k_T^\nu a_{3T}^{\perp g} + P^\mu N^\nu a_{4T}^{\perp g} + N^\mu P^\nu a_{5T}^{\perp g} \\
    &\qquad\qquad\qquad\qquad + P^\mu k_T^\nu a_{6T}^{\perp g} + k_T^\mu P^\nu a_{7T}^{\perp g} + N^\mu k_T^\nu a_{8T}^{\perp g} + k_T^\mu N^\nu a_{9T}^{\perp g} + M^2g_T^{\mu\nu}a_{0T}^{\perp g}\Bigr] \\
    &\qquad
    - M\Bigl[P^\mu\epsilon_T^{\nu S_T}a_{1T}^g + P^\nu\epsilon_T^{\mu S_T}a_{2T}^g + N^\mu\epsilon_T^{\nu S_T}a_{3T}^g \\
    &\qquad\qquad\qquad\qquad + N^\nu\epsilon_T^{\mu S_T}a_{4T}^g + k_T^\mu\epsilon_T^{\nu S_T}a_{5T}^g + k_T^\nu\epsilon_T^{\mu S_T}a_{6T}^g\Bigr] \\
    &\qquad - \lambda\Bigl[P^\mu\epsilon_T^{\nu k_T}a_{1L}^g + P^\nu\epsilon_T^{\mu k_T}a_{2L}^g + N^\mu\epsilon_T^{\nu k_T}a_{3L}^g \\
    &\qquad\qquad\qquad\qquad + N^\nu\epsilon_T^{\mu k_T}a_{4L}^g + k_T^\mu\epsilon_T^{\nu k_T}a_{5L}^g + k_T^\nu\epsilon_T^{\mu k_T}a_{6L}^g\Bigr]\biggr\}\,,
\end{aligned}
\label{eq:gluon-gravitational-TMD-parametrization}
\end{equation}
where the dependence on the Wilson-line direction is made explicit through the rescaling-invariant lightlike vector~\cite{Lorce:2015lna, Lorce:2023zzg}
\begin{equation}
    N^\mu = \frac{M^2 n^\mu}{P\cdot n}\,, \qquad N^2 = 0\,, \qquad P\cdot N = M^2\,.
    \label{eq:gluon-rescaling-invariant-vector}
\end{equation}
Here, $M$ is the hadron mass.  The tensor
\begin{equation}
    g_T^{\mu\nu} = g^{\mu\nu} - \frac{P^\mu N^\nu + P^\nu N^\mu}{M^2} + \frac{N^\mu N^\nu}{M^2}
    \label{eq:gluon-transverse-projector}
\end{equation}
projects onto the transverse subspace orthogonal to $P^\mu$ and $N^\mu$.  Accordingly, the transverse gluon momentum and the decomposition of the target polarization are given by
\begin{equation}
    k_T^\mu = g_T^{\mu\nu}k_\nu\,, \qquad k_T^2 = - \bm{k}_T^2\,, \qquad S^\mu = \frac{\lambda}{M}(P^\mu - N^\mu) + S_T^\mu\,,
    \label{eq:gluon-transverse-momentum-spin}
\end{equation}
where $\lambda$ and $S_T^\mu$ denote the longitudinal and transverse polarizations, respectively.  The transverse Levi-Civita tensor is defined as
\begin{equation}
    \epsilon_T^{\mu\nu} = \frac{\epsilon^{\mu\nu\alpha\beta}N_\alpha P_\beta}{M^2}\,, \qquad \epsilon_{0123} = +1\,, \qquad \epsilon_T^{12} = +1\,,
    \label{eq:gluon-transverse-epsilon}
\end{equation}
and the contractions in Eq.~\eqref{eq:gluon-gravitational-TMD-parametrization} use the shorthand notation
\begin{equation}
    \epsilon_T^{\mu v_T} \coloneqq \epsilon_T^{\mu\nu}v_{T\nu}\,, \qquad \epsilon_T^{k_TS_T} \coloneqq \epsilon_T^{\alpha\beta}k_{T\alpha}S_{T\beta}\,.
\end{equation}

The scalar coefficients in Eq.~\eqref{eq:gluon-gravitational-TMD-parametrization} are referred to as gluon gravitational TMDs~\cite{Lorce:2023zzg}.  Relations between these functions and ordinary TMDs will be discussed below.

\subsection{Relation between TMDs and gravitational TMDs for gluons}

To make the connection with gluon TMDs explicit, we introduce the field-strength correlator~\cite{Lorce:2013pza}
\begin{equation}
    \Phi_g^{\alpha\beta; \rho\sigma} = \frac{1}{xP^+}\int\frac{dz^-d^2\bm{z}_T}{(2\pi)^3}\,e^{ik\cdot z}\mel{P, S}{F_a^{\alpha\beta}(- z/2)\mathcal{U}_\eta^{ab}F_b^{\rho\sigma}(z/2)}{P, S}\eval_{z^+ = 0}\,.
    \label{eq:gluon-FF-TMD}
\end{equation}
The adjoint staple $\mathcal{U}_\eta$ corresponds to the WW gauge-link structure, with both fundamental staple links passing through the same light-front infinity.  This differs from the dipole gauge-link structure, in which the two staple links pass through opposite light-front infinities~\cite{Dominguez:2010xd, Dominguez:2011wm}.  Throughout this work, we adopt the WW definition for the gluon TMDs entering the gic EMT.

Using $D_{\mathrm{pure}}^+A_{\mathrm{phys}}^\alpha = F^{+\alpha}$, together with translation invariance of the forward matrix element and integration by parts for the pure-gauge Wilson-line prescription specified above, the transverse-momentum-dependent EMT can be expressed as
\begin{equation}
    \mathcal{T}_g^{\mu\nu} = - k^\nu g_{\alpha\beta}\Re\Phi_g^{\mu\alpha; +\beta} + \frac{xP^+}{4}g^{\mu\nu}g_{\alpha\rho}g_{\beta\sigma}\Phi_g^{\alpha\beta; \rho\sigma}\,, \qquad \nu = +, 1, 2\,.
    \label{eq:gluon-canonical-TMD-k-factor}
\end{equation}
The real part in the first term follows from the Hermitian ordering of the bilocal EMT, whereas the complete Lorentz contraction in the second term is already real by hermiticity.

For $\nu = -$, the momentum factor must instead be retained inside the $k^-$ integral,
\begin{equation}
    \mathcal{T}_g^{\mu-} = - g_{\alpha\beta}\int dk^-\,k^-\Re\widehat{\Phi}_g^{\mu\alpha; +\beta}(P, k, S; \eta) + \frac{xP^+}{4}g^{\mu-}g_{\alpha\rho}g_{\beta\sigma}\Phi_g^{\alpha\beta; \rho\sigma}\,,
    \label{eq:gluon-canonical-TMD-minus}
\end{equation}
where $\widehat{\Phi}_g^{\alpha\beta; \rho\sigma}$ is the fully unintegrated counterpart of Eq.~\eqref{eq:gluon-FF-TMD}.

The first term in Eq.~\eqref{eq:gluon-canonical-TMD-k-factor} has the same structure as the quark result of Ref.~\cite{Lorce:2023zzg} and encodes the densities and fluxes of canonical gluon momentum. The additional metric term originates from the $- g^{\mu\nu}\mathcal{L}_g$ contribution to the gluon EMT and is also present in the kinetic construction of Ref.~\cite{Saha:2026veh}.  It vanishes for $\mathcal{T}_g^{++}$, $\mathcal{T}_g^{+i}$, and $\mathcal{T}_g^{i+}$, but contributes to the light-front energy density $\mathcal{T}_g^{+-}$ and provides an additional isotropic contribution to the transverse stress $\mathcal{T}_g^{ij}$, without directly affecting its traceless shear part.

Equation~\eqref{eq:gluon-canonical-TMD-k-factor} shows that the components with $\nu = +, 1, 2$ depend only on the contractions $- g_{\alpha\beta}\Re\Phi_g^{\mu\alpha; +\beta}$ and $g_{\alpha\rho}g_{\beta\sigma}\Phi_g^{\alpha\beta; \rho\sigma}$.  We therefore restrict the following parametrization to the projections entering these contractions and organize them according to their twist, in analogy with the quark analysis of Ref.~\cite{Lorce:2023zzg}.

For a spin-$1/2$ target, the $\mu = +$ projection involves the twist-two transverse trace,
\begin{equation}
    - g_{\alpha\beta}\Re\Phi_g^{+\alpha; +\beta} = - g_{Tij}\Phi_g^{+i; +j} = f_1^g - \frac{\epsilon_T^{k_TS_T}}{M}f_{1T}^{\perp g}\,,
    \label{eq:gluon-standard-leading-trace}
\end{equation}
where $f_1^g$ and $f_{1T}^{\perp g}$ denote the unpolarized and Sivers gluon TMDs, respectively~\cite{Mulders:2000sh, Lorce:2013pza}.

For $\mu = i$, the contraction $- g_{\alpha\beta}\Re\Phi_g^{i\alpha; +\beta}$ is the sum of two twist-three projections, $\Re\Phi_g^{+i; +-}$ and $g_{Tjl}\Re\Phi_g^{ij; l+}$.  Using the gluon TMD notation of Ref.~\cite{Lorce:2013pza}, these projections are parametrized as
\begin{align}
    \Re\Phi_g^{+i; +-} &= \frac{M}{P^+}\left(\frac{k_T^i}{M}f^{\perp g} - \epsilon_T^{iS_T}f_T^g - \lambda\frac{\epsilon_T^{ik_T}}{M}f_L^{\perp g} + \frac{K_T^{ij}}{M^2}\epsilon_{TjS_T}f_T^{\perp g}\right)\,,
    \label{eq:gluon-twist3-electric} \\
    g_{Tjl}\Re\Phi_g^{ij; l+} &= \frac{M}{P^+}\left(- \frac{k_T^i}{M}\bar{g}^{\perp g} - \epsilon_T^{iS_T}\bar{g}_T^g - \lambda\frac{\epsilon_T^{ik_T}}{M}\bar{g}_L^{\perp g} - \frac{K_T^{ij}}{M^2}\epsilon_{TjS_T}\bar{g}_T^{\perp g}\right)\,,
    \label{eq:gluon-twist3-magnetic}
\end{align}
where $K_T^{ij} = k_T^ik_T^j - \frac{1}{2}k_T^2g_T^{ij}$ is the symmetric traceless transverse tensor, with $k_T^2 = - \bm{k}_T^2$, and $\epsilon_{TjS_T} = g_{Tj\alpha}\epsilon_T^{\alpha S_T}$.  The bar on $\bar{g}$ distinguishes a separate set of real distributions and does not denote complex conjugation.

The remaining $\mu = -$ projection and the complete Lorentz contraction involve twist-four correlators.  For a spin-$1/2$ target, the parity-even scalar structures allowed in these projections are $1$ and $\epsilon_T^{k_TS_T} / M$.  We therefore define
\begin{align}
    - g_{\alpha\beta}\Re\Phi_g^{-\alpha; +\beta} &= \left(\frac{M}{P^+}\right)^2\left(\mathsf{f}_3^g - \frac{\epsilon_T^{k_TS_T}}{M}\mathsf{f}_{3T}^{\perp g}\right)\,,
    \label{eq:gluon-twist4-vector-projection} \\
    \frac{x}{4}g_{\alpha\rho}g_{\beta\sigma}\Phi_g^{\alpha\beta; \rho\sigma} &= \left(\frac{M}{P^+}\right)^2\left(s_g - \frac{\epsilon_T^{k_TS_T}}{M}s_{gT}^{\perp}\right)\,.
    \label{eq:gluon-twist4-scalar-projection}
\end{align}
Here, $\mathsf{f}_3^g$, $\mathsf{f}_{3T}^{\perp g}$, $s_g$, and $s_{gT}^{\perp}$ are introduced in this work as the scalar coefficients of the specific twist-four combinations selected by the gluon EMT.  The functions $\mathsf{f}_3^g$ and $s_g$ describe the polarization-independent contributions, whereas $\mathsf{f}_{3T}^{\perp g}$ and $s_{gT}^{\perp}$ multiply the transverse-spin-dependent structure.  The notation for the first pair follows the analogy with the twist-four quark vector correlator in Ref.~\cite{Lorce:2023zzg}; it does not imply an identification with individually named gluon TMDs in a complete twist-four basis.

All functions displayed above are real functions of $(x, \bm{k}_T^2; \eta)$, with these arguments suppressed for brevity.  The omission of other projections does not imply that the corresponding gluon TMDs vanish.  Moreover, the $\nu = -$ components involve first $k^-$ moments of the fully unintegrated correlator, which are not determined by the ordinary TMD correlator alone.

To express the gravitational TMDs in terms of the projections introduced above, we define the combinations
\begin{equation}
\begin{split}
    &\mathsf{f}_1^g \coloneqq f_1^g\,, \qquad \mathsf{f}_{1T}^{\perp g} \coloneqq f_{1T}^{\perp g}\,, \qquad \mathsf{f}^{\perp g} \coloneqq f^{\perp g} - \bar{g}^{\perp g}\,, \\
    &\mathsf{f}_T^g \coloneqq f_T^g + \bar{g}_T^g\,, \qquad \mathsf{f}_L^{\perp g} \coloneqq f_L^{\perp g} + \bar{g}_L^{\perp g}\,, \qquad \mathsf{f}_T^{\perp g} \coloneqq - \left(f_T^{\perp g} - \bar{g}_T^{\perp g}\right)\,.
    \label{eq:gluon-EMT-TMD-combinations}
\end{split}
\end{equation}
These definitions provide shorthand notation for the twist-two and twist-three combinations selected by the gluon EMT, rather than introducing additional independent TMDs.

We also introduce
\begin{equation}
    \mathsf{f}_T^{+g} \coloneqq \mathsf{f}_T^g + \frac{\bm{k}_T^2}{2M^2}\mathsf{f}_T^{\perp g}\,, \qquad \widetilde{P}^\mu \coloneqq P^\mu - \frac{1}{2}N^\mu = \left[P^+, 0, \bm{0}_T\right]\,.
    \label{eq:gluon-TMD-auxiliary-definitions}
\end{equation}
Using the identity
\begin{equation}
    K_T^{ij}\epsilon_{TjS_T} = k_T^i\epsilon_T^{k_TS_T} + \frac{\bm{k}_T^2}{2}\epsilon_T^{iS_T}\,,
\end{equation}
the vector contraction entering Eq.~\eqref{eq:gluon-canonical-TMD-k-factor} can be written as
\begin{equation}
\begin{split}
    - g_{\alpha\beta}\Re\Phi_g^{\mu\alpha; +\beta} &= \frac{1}{P^+}\left[\widetilde{P}^\mu\mathsf{f}_1^g + k_T^\mu\mathsf{f}^{\perp g} + N^\mu\mathsf{f}_3^g - \frac{\epsilon_T^{k_TS_T}}{M}\left(\widetilde{P}^\mu\mathsf{f}_{1T}^{\perp g} + k_T^\mu\mathsf{f}_T^{\perp g} + N^\mu\mathsf{f}_{3T}^{\perp g}\right) - M\epsilon_T^{\mu S_T}\mathsf{f}_T^{+g} - \lambda\epsilon_T^{\mu k_T}\mathsf{f}_L^{\perp g}\right]\,.
\end{split}
\label{eq:gluon-contracted-vector-parametrization}
\end{equation}
This expression combines the twist-two, twist-three, and twist-four vector projections in a form suitable for comparison with Eq.~\eqref{eq:gluon-gravitational-TMD-parametrization}.

Setting $\nu = +$ in Eq.~\eqref{eq:gluon-canonical-TMD-k-factor}, and retaining the scalar contribution defined in Eq.~\eqref{eq:gluon-twist4-scalar-projection}, gives
\begin{equation}
\begin{split}
    &a_1^g = x\mathsf{f}_1^g\,, \qquad a_{1T}^{\perp g} = x\mathsf{f}_{1T}^{\perp g}\,, \qquad \frac{1}{2}a_1^g + a_5^g = x\mathsf{f}_3^g + s_g\,, \qquad \frac{1}{2}a_{1T}^{\perp g} + a_{5T}^{\perp g} = x\mathsf{f}_{3T}^{\perp g} + s_{gT}^{\perp}\,, \\
    &a_7^g = x\mathsf{f}^{\perp g}\,, \qquad a_{7T}^{\perp g} = x\mathsf{f}_T^{\perp g}\,, \qquad a_{2L}^g = x\mathsf{f}_L^{\perp g}\,, \qquad a_{2T}^g = x\mathsf{f}_T^{+g}\,.
    \label{eq:gluon-matching-plus-column}
\end{split}
\end{equation}
Similarly, the components with $\nu = i$ yield
\begin{equation}
\begin{split}
    &a_3^g = \mathsf{f}^{\perp g}\,, \qquad a_{3T}^{\perp g} = \mathsf{f}_T^{\perp g}\,, \qquad a_6^g = \mathsf{f}_1^g\,, \qquad a_{6T}^{\perp g} = \mathsf{f}_{1T}^{\perp g}\,, \qquad \frac{1}{2}a_6^g + a_8^g = \mathsf{f}_3^g\,, \qquad \frac{1}{2}a_{6T}^{\perp g} + a_{8T}^{\perp g} = \mathsf{f}_{3T}^{\perp g}\,, \\
    &a_{6L}^g = \mathsf{f}_L^{\perp g}\,, \qquad a_{6T}^g = \mathsf{f}_T^{+g}\,, \qquad a_0^g = s_g\,, \qquad a_{0T}^{\perp g} = s_{gT}^{\perp}\,, \qquad a_{1L}^g = a_{3L}^g = a_{5L}^g = 0\,, \qquad a_{1T}^g = a_{3T}^g = a_{5T}^g = 0\,.
    \label{eq:gluon-matching-transverse-column}
\end{split}
\end{equation}
The factors of $1/2$ in these relations originate from the decomposition $\widetilde{P}^\mu = P^\mu - N^\mu / 2$.  In contrast to the quark result of Ref.~\cite{Lorce:2023zzg}, the functions $a_0^g$ and $a_{0T}^{\perp g}$ need not vanish: they encode the additional metric term in the gluon EMT.  The same term contributes to $a_5^g$ and $a_{5T}^{\perp g}$ through the $\mathcal{T}_g^{-+}$ component.

The remaining coefficients enter the $\nu = -$ column and require information beyond the ordinary TMD correlator.  To display this dependence explicitly, we define the normalized first $k^-$ moment
\begin{equation}
    \check{\Phi}_g^\mu \coloneqq - \frac{P^+}{M^2}g_{\alpha\beta}\int dk^-\,k^-\Re\widehat{\Phi}_g^{\mu\alpha; +\beta}(P, k, S; \eta)\,.
    \label{eq:gluon-first-kminus-moment}
\end{equation}
Equation~\eqref{eq:gluon-canonical-TMD-minus} then becomes
\begin{equation}
    \mathcal{T}_g^{\mu-} = \frac{M^2}{P^+}\left[\check{\Phi}_g^\mu + g^{\mu-}\left(s_g - \frac{\epsilon_T^{k_TS_T}}{M}s_{gT}^{\perp}\right)\right]\,.
    \label{eq:gluon-minus-column-moment}
\end{equation}
We parametrize $\check{\Phi}_g^\mu$ by the same Lorentz structures as the right-hand side of Eq.~\eqref{eq:gluon-contracted-vector-parametrization}, replacing each $\mathsf{f}$ by $\check{\mathsf{f}}$.  Comparing with Eq.~\eqref{eq:gluon-gravitational-TMD-parametrization}, we obtain
\begin{equation}
\begin{split}
    &a_4^g = \check{\mathsf{f}}_1^g - \frac{x}{2}\mathsf{f}_1^g + s_g\,, \qquad a_{4T}^{\perp g} = \check{\mathsf{f}}_{1T}^{\perp g} - \frac{x}{2}\mathsf{f}_{1T}^{\perp g} + s_{gT}^{\perp}\,, \qquad a_9^g = \check{\mathsf{f}}^{\perp g} - \frac{x}{2}\mathsf{f}^{\perp g}\,, \qquad a_{9T}^{\perp g} = \check{\mathsf{f}}_T^{\perp g} - \frac{x}{2}\mathsf{f}_T^{\perp g}\,, \\
    &a_2^g + \frac{1}{2}a_4^g = \check{\mathsf{f}}_3^g - \frac{x}{2}\mathsf{f}_3^g - \frac{1}{2}s_g\,, \qquad a_{2T}^{\perp g} + \frac{1}{2}a_{4T}^{\perp g} = \check{\mathsf{f}}_{3T}^{\perp g} - \frac{x}{2}\mathsf{f}_{3T}^{\perp g} - \frac{1}{2}s_{gT}^{\perp}\,, \qquad a_{4L}^g = \check{\mathsf{f}}_L^{\perp g} - \frac{x}{2}\mathsf{f}_L^{\perp g}\,, \qquad a_{4T}^g = \check{\mathsf{f}}_T^{+g} - \frac{x}{2}\mathsf{f}_T^{+g}\,.
    \label{eq:gluon-minus-column-coefficients}
\end{split}
\end{equation}
The checked functions play the same role as their quark counterparts introduced in Eqs.~(25) and (26) of Ref.~\cite{Lorce:2023zzg}: they parametrize the information in the $\nu = -$ components that is not fixed by the ordinary TMD projections.  In the present construction, these functions are defined through the first $k^-$ moment in Eq.~\eqref{eq:gluon-first-kminus-moment}, with the scalar contribution to the gluon EMT kept separate.

In the following sections, we restrict our analysis of mechanical properties and small-$x$ distributions to the components $\mathcal{T}^{++}$, $\mathcal{T}^{+i}$, $\mathcal{T}^{i+}$, and $\mathcal{T}^{ij}$, with $i,j = 1,2$.  The $\nu = -$ components and the associated checked functions are therefore not needed for the subsequent analysis.

\section{Mechanical properties}
\label{sec:gluon-mechanical-properties}

The momentum-space mechanical interpretation of gravitational TMDs has been discussed for quarks in Ref.~\cite{Lorce:2023zzg} and for gluons using the kinetic (Belinfante--Rosenfeld) EMT in Ref.~\cite{Saha:2026veh}.  To prepare for the small-$x$ analysis, we summarize the relevant momentum densities, transverse flux of inertia, and transverse stress for the light-front gic gluon EMT adopted here.  The discussion follows the quark analysis, highlighting the gluon TMD combinations and the additional scalar contribution where they modify the corresponding relations.

\subsection{Densities of longitudinal and transverse momentum}
\label{subsec:gluon-momentum-densities}

The longitudinal and transverse canonical momentum densities are
\begin{equation}
\begin{split}
    \mathcal{T}_g^{++} &= xP^+\left(f_1^g - \frac{\epsilon_T^{k_TS_T}}{M}f_{1T}^{\perp g}\right)\,, \\
    \mathcal{T}_g^{+i} &= k_T^i\left(f_1^g - \frac{\epsilon_T^{k_TS_T}}{M}f_{1T}^{\perp g}\right)\,, \qquad i = 1, 2\,.
\end{split}
\label{eq:gluon-momentum-densities}
\end{equation}
These expressions have the same structure as their quark counterparts and involve only the unpolarized and Sivers TMDs.  The scalar contribution is absent because $g^{++} = g^{+i} = 0$.  In particular, the transverse momentum density is directed along $\bm{k}_T$, with the Sivers term introducing a polarization-dependent angular modulation of its magnitude.

\subsection{Transverse flux of inertia}
\label{subsec:gluon-inertia-flux}

In light-front kinematics, longitudinal momentum plays the role of inertia under transverse boosts.  The component $\mathcal{T}_g^{i+}$ therefore describes the transverse flux of inertia~\cite{Lorce:2023zzg},
\begin{equation}
    \mathcal{T}_g^{i+} = x\left(\mathsf{f}^{\perp g} - \frac{\epsilon_T^{k_TS_T}}{M}\mathsf{f}_T^{\perp g}\right)k_T^i - xM\epsilon_T^{iS_T}\mathsf{f}_T^{+g} - x\lambda\epsilon_T^{ik_T}\mathsf{f}_L^{\perp g}\,.
    \label{eq:gluon-transverse-inertia-flux}
\end{equation}
As in the quark case, this flux need not coincide with the canonical momentum density $\mathcal{T}_g^{+i}$.  The relevant twist-three functions are now the gluon combinations defined above; in particular, $\mathsf{f}^{\perp g} = f^{\perp g} - \bar{g}^{\perp g}$.  The scalar contribution is again absent since $g^{i+} = 0$.

For the same gauge-link choice and correlator normalization, $\mathcal{T}_g^{++}$ and $\mathcal{T}_g^{i+}$ coincide with the corresponding real projections of the kinetic construction in Ref.~\cite{Saha:2026veh}.  The distinction lies in the transverse momentum density: the canonical expression in Eq.~\eqref{eq:gluon-momentum-densities} is determined by the twist-two trace, whereas the kinetic component $\mathcal{T}_{g, \mathrm{gik}}^{+i}$ involves twist-three correlators.

Following Ref.~\cite{Lorce:2023zzg}, we define an effective transverse velocity by
\begin{equation}
    v_{g, T}^i \coloneqq \frac{\mathcal{T}_g^{i+}}{\mathcal{T}_g^{++}}\, \qquad v_{g, T}^i - \frac{k_T^i}{xP^+} = \frac{\mathcal{T}_g^{i+} - \mathcal{T}_g^{+i}}{\mathcal{T}_g^{++}}\,,
    \label{eq:gluon-effective-transverse-velocity}
\end{equation}
where $x \neq 0$ and $\mathcal{T}_g^{++} \neq 0$.  The second relation follows from Eq.~\eqref{eq:gluon-momentum-densities} and shows that $v_{g, T}^i = k_T^i/(xP^+)$ precisely when the transverse momentum density equals the transverse flux of inertia.  In general, the two mixed components differ, and the polarization-dependent terms in Eq.~\eqref{eq:gluon-transverse-inertia-flux} can also change the direction of the velocity relative to $\bm{k}_T$.

At generic nonzero $x$ and $\bm{k}_T$, requiring $\mathcal{T}_g^{i+} = \mathcal{T}_g^{+i}$ for arbitrary target polarization gives
\begin{equation}
    x\mathsf{f}^{\perp g} = f_1^g\,, \qquad x\mathsf{f}_T^{\perp g} = f_{1T}^{\perp g}\,, \qquad \mathsf{f}_T^{+g} = \mathsf{f}_L^{\perp g} = 0\,.
    \label{eq:gluon-mixed-component-symmetry}
\end{equation}
These conditions concern only the mixed components considered here and do not imply symmetry of the full EMT.  None of these relations is assumed in the present analysis.

For later comparison, we introduce the ratio of the polarization-independent coefficients,
\begin{equation}
    R_v^g \coloneqq \frac{x\mathsf{f}^{\perp g}}{f_1^g} = \frac{x\left(f^{\perp g} - \bar{g}^{\perp g}\right)}{f_1^g}\,,
    \label{eq:gluon-inertia-flux-ratio}
\end{equation}
where $f_1^g \neq 0$.  The condition $R_v^g = 1$ tests the first relation in Eq.~\eqref{eq:gluon-mixed-component-symmetry}.  Its quark counterpart, $xf^{\perp q} = f_1^q$, follows from the free-quark equation of motion~\cite{Lorce:2014hxa}, as noted in Ref.~\cite{Lorce:2023zzg}, motivating the use of $R_v^g = 1$ as a reference for comparison with free partons.  This condition alone, however, neither establishes free gluon dynamics nor guarantees equality of the mixed components for arbitrary target polarization.

\subsection{Transverse pressure and shear forces}
\label{subsec:gluon-pressure-shear}

The transverse stress is related to the inertia flux by
\begin{equation}
    \mathcal{T}_g^{ij} = \frac{k_T^j}{xP^+}\mathcal{T}_g^{i+} + \frac{M^2}{P^+}g_T^{ij}\left(s_g - \frac{\epsilon_T^{k_TS_T}}{M}s_{gT}^{\perp}\right)\,.
    \label{eq:gluon-stress-flux-relation}
\end{equation}
The second term is the additional scalar contribution absent from the quark relation.  For $\bm{k}_T \neq 0$, we use the stress decomposition of Ref.~\cite{Lorce:2023zzg},
\begin{equation}
    \mathcal{T}_g^{ij} = - g_T^{ij}\sigma_g + \left(\frac{1}{2}g_T^{ij} - \frac{k_T^ik_T^j}{k_T^2}\right)\Pi_g + \frac{k_T^i\epsilon_T^{jk_T} + k_T^j\epsilon_T^{ik_T}}{2k_T^2}\Pi_g^S + \epsilon_T^{ij}\Pi_g^A\,.
    \label{eq:gluon-stress-decomposition}
\end{equation}
Here, $\sigma_g$ is the isotropic transverse stress, $\Pi_g$ is the normal-stress anisotropy, and $\Pi_g^S$ and $\Pi_g^A$ describe the remaining symmetric and antisymmetric stresses, respectively.

Introducing $\mathsf{f}_T^{-g} \coloneqq \mathsf{f}_T^g - \bm{k}_T^2\mathsf{f}_T^{\perp g}/(2M^2)$, we obtain
\begin{equation}
\begin{split}
    \Pi_g &= - \frac{1}{P^+}\left(k_T^2\mathsf{f}^{\perp g} - M\epsilon_T^{k_TS_T}\mathsf{f}_T^{-g}\right)\,, \\
    \sigma_g &= \frac{1}{2}\Pi_g - \frac{M^2}{P^+}\left(s_g - \frac{\epsilon_T^{k_TS_T}}{M}s_{gT}^{\perp}\right)\,, \\
    \Pi_g^A = \frac{1}{2}\Pi_g^S &= - \frac{1}{2P^+}\left[\lambda k_T^2\mathsf{f}_L^{\perp g} + M(k_T\cdot S_T)\mathsf{f}_T^{+g}\right]\,.
\end{split}
\label{eq:gluon-pressure-shear-TMDs}
\end{equation}
The scalar metric term, also present in the kinetic gluon EMT of Ref.~\cite{Saha:2026veh}, shifts the isotropic stress without contributing to the traceless symmetric or antisymmetric stress.  The stress--flux relation in Eq.~\eqref{eq:gluon-stress-flux-relation}, however, relies on the canonical operator structure and does not hold in general for the kinetic EMT.  Consequently, the quark relation $\sigma_q = \Pi_q/2$ is generally lost, whereas $\Pi_g^A = \Pi_g^S/2$ remains valid.  Unlike the quark transverse stress, which is determined by twist-three vector TMDs, the complete gluon transverse stress also requires the twist-four scalar projection.  Both contributions in Eq.~\eqref{eq:gluon-stress-flux-relation} scale as $1/P^+$ at fixed $M$ and $\bm{k}_T$.

For an unpolarized target, recovering the free-particle reference $\mathcal{T}_{g,U,\mathrm{free}}^{ij} = k_T^ik_T^jf_1^g/(xP^+)$ requires both $R_v^g = 1$ and $s_g = 0$.

\section{Small-$x$ gravitational TMDs in the eikonal approximation}
\label{sec:smallx-eikonal}

We now apply the preceding relations to gluons and sea quarks at small $x$.  We first identify the gluon EMT components that survive in the eikonal approximation and then express the relevant distributions in terms of Wilson-line correlators.  For the numerical analysis, a common fundamental dipole is evolved from an MV initial condition using fixed-coupling LO BK evolution.  The sea-quark inertia flux is considered separately through the high-$k_T$ matching of the quark TMDs.

\subsection{Gluon TMDs in the eikonal approximation}
\label{subsec:smallx-gluon-emt}

Within the CGC framework, the fast partons are represented by color sources that are frozen in light-front time and generate a classical small-$x$ gluon field~\cite{McLerran:1993ka, Iancu:2000hn}.  For a target moving with large $P^+$, we adopt the corresponding classical solution in covariant gauge,
\begin{equation}
    A^\mu(z) = \delta^{\mu+}\alpha(z^-, \bm{z}_T)\,,
    \label{eq:sx-eikonal-background}
\end{equation}
where $\alpha = \alpha_at^a$ is localized in the longitudinal direction.  We retain a finite longitudinal profile to preserve the ordering of color matrices in the Wilson lines.  For a longitudinal extent $L^-$ satisfying $xP^+L^- \ll 1$, however, the Fourier phase across the target is neglected, while the full transverse dependence and all orders in the background field are retained through Wilson lines.  Together with the field configuration above, this implements the shockwave approximation at eikonal accuracy~\cite{Altinoluk:2024zom}.  Hereafter, $|_{\mathrm{eik}}$ denotes this approximation at leading classical order.

The only nonzero field-strength components are
\begin{equation}
    F^{+i} = - \partial^i\alpha\,, \qquad F^{+-} = F^{ij} = F^{-i} = 0\,.
    \label{eq:sx-eikonal-field-strengths}
\end{equation}
Consequently, among the vector contractions $- g_{\alpha\beta}\Re\Phi_g^{\mu\alpha; +\beta}$ entering the canonical EMT, only the $\mu = +$ projection can survive.  The complete scalar contraction also vanishes, since a nonzero $F^{+i}$ would have to contract with $F_{+i} = g_{Tij}F^{-j} = 0$.  Using Eqs.~\eqref{eq:gluon-twist3-electric}--\eqref{eq:gluon-twist4-scalar-projection}, we therefore obtain
\begin{equation}
\begin{split}
    \left.\mathsf{f}^{\perp g}\right|_{\mathrm{eik}} &= \left.\mathsf{f}_T^g\right|_{\mathrm{eik}} = \left.\mathsf{f}_L^{\perp g}\right|_{\mathrm{eik}} = \left.\mathsf{f}_T^{\perp g}\right|_{\mathrm{eik}} = 0\,, \\
    \left.\mathsf{f}_3^g\right|_{\mathrm{eik}} &= \left.\mathsf{f}_{3T}^{\perp g}\right|_{\mathrm{eik}} = \left.s_g\right|_{\mathrm{eik}} = \left.s_{gT}^{\perp}\right|_{\mathrm{eik}} = 0\,.
\end{split}
\label{eq:sx-vanishing-gluon-TMD-combinations}
\end{equation}

The surviving transverse trace contains $f_1^g$ and $f_{1T}^{\perp g}$ before the small-$x$ matching.  For the same-direction gauge links adopted here, the leading Wilson-line expression is symmetric under interchange of its two transverse endpoints, as made explicit in Sec.~\ref{subsec:smallx-Wilson-inputs}.  Its Fourier transform is therefore even under $\bm{k}_T \to -\bm{k}_T$ at fixed target polarization.  Since the Sivers structure is odd under this transformation~\cite{Boer:2015pni, Yao:2018vcg},
\begin{equation}
    \left.xf_{1T}^{\perp g,\mathrm{WW}}\right|_{\mathrm{eik}} = 0\,.
    \label{eq:sx-WW-Sivers-zero}
\end{equation}
Thus, the only distribution entering the gluon EMT at this order is the unpolarized WW distribution.

The resulting EMT components relevant to the mechanical discussion reduce to
\begin{equation}
\begin{split}
    \left.\mathcal{T}_g^{++}\right|_{\mathrm{eik}} = \left.xP^+f_1^{g,\mathrm{WW}}\right|_{\mathrm{eik}}\,, \qquad \left.\mathcal{T}_g^{+i}\right|_{\mathrm{eik}} = \left.k_T^if_1^{g,\mathrm{WW}}\right|_{\mathrm{eik}}\,, \qquad \left.\mathcal{T}_g^{i+}\right|_{\mathrm{eik}} = \left.\mathcal{T}_g^{ij}\right|_{\mathrm{eik}} = 0\,.
\end{split}
\label{eq:sx-gluon-eikonal-components}
\end{equation}
The canonical transverse momentum density remains nonzero, whereas the transverse flux of inertia vanishes in the eikonal approximation.  This difference holds at fixed transverse momentum and is not a cancellation arising from angular integration.  Equations~\eqref{eq:gluon-effective-transverse-velocity}, \eqref{eq:gluon-inertia-flux-ratio}, and \eqref{eq:gluon-pressure-shear-TMDs} give
\begin{equation}
    \left.R_v^g\right|_{\mathrm{eik}} = 0\,, \qquad \left.v_{g, T}^i\right|_{\mathrm{eik}} = 0\,, \qquad \left.\sigma_g\right|_{\mathrm{eik}} = \left.\Pi_g\right|_{\mathrm{eik}} = \left.\Pi_g^S\right|_{\mathrm{eik}} = \left.\Pi_g^A\right|_{\mathrm{eik}} = 0\,.
    \label{eq:sx-gluon-mechanical-properties}
\end{equation}
The condition $R_v^g = 1$ is necessary for a free-particle interpretation of these distributions.  The result $R_v^g = 0$ is stronger than a mere violation of this condition: the nonzero canonical transverse momentum density has no accompanying transverse flux of inertia at this order.  Consequently, the entire transverse momentum density remains in the difference $\mathcal{T}_g^{+i} - \mathcal{T}_g^{i+}$, rather than being associated with the transverse transport expected for free on-shell particles.

\subsection{WW gluon and sea-quark distributions}
\label{subsec:smallx-Wilson-inputs}

To compare gluons and sea quarks, we restrict the remainder of this section, including the high-$k_T$ analysis below, to an unpolarized target.  The relevant quark vector projections therefore reduce to~\cite{Lorce:2023zzg}
\begin{equation}
    \Phi_{q}^{[\gamma^+]} = f_1^q\,, \qquad \Phi_{q}^{[\gamma^i]} = \frac{k_T^i}{P^+}f^{\perp q}\,.
    \label{eq:sx-spin-averaged-quark-projections}
\end{equation}
This projection does not assume that the spin-dependent TMDs vanish for a polarized target.

We now express the surviving gluon distribution and the sea-quark input in terms of Wilson-line correlators.  The small-$x$ matching formulas below are understood in the eikonal approximation, and the corresponding restriction is suppressed for brevity.  For definiteness, we choose future-pointing staples, $\eta = +1$, and introduce the fundamental Wilson lines
\begin{equation}
    U_{\bm{b}} \coloneqq L_{\bm{b}}(-\infty, +\infty)\,, \qquad L_{\bm{b}}(a^-, b^-) \coloneqq \mathcal{P}\exp\left[ig\int_{b^-}^{a^-}d\zeta^-\,\alpha(\zeta^-, \bm{b})\right]\,.
    \label{eq:sx-longitudinal-lines}
\end{equation}
Here, $L_{\bm{b}}(a^-, b^-)$ transports color from $b^-$ to $a^-$ at fixed transverse position $\bm{b}$.  Transverse connections at infinity are understood and reduce to unity in the gauge of Eq.~\eqref{eq:sx-eikonal-background}.  The notation $\expval{\cdots}_{Y}$ denotes averaging over the unpolarized target ensemble at rapidity $Y = \ln(x_0 / x)$.

The WW distribution involves the operator
\begin{equation}
    \mathcal{O}_{\mathrm{WW}}(\bm{b}_1, \bm{b}_2) \coloneqq - \frac{1}{N_c}\sum_{i = 1}^{2}\Re\Tr\left[(\partial_iU_{\bm{b}_1})U_{\bm{b}_2}^\dagger(\partial_iU_{\bm{b}_2})U_{\bm{b}_1}^\dagger\right]\,,
    \label{eq:sx-WW-operator}
\end{equation}
where $\partial_i$ denotes differentiation with respect to the Euclidean transverse coordinate.  In the normalization of Eq.~\eqref{eq:gluon-FF-TMD}, the matching reads~\cite{Dominguez:2011wm, Hagiwara:2024wqz}
\begin{equation}
    xf_1^{g, \mathrm{WW}}(x, \bm{k}_T^2) = \frac{2N_c}{\alpha_s}\int\frac{d^2\bm{b}_1d^2\bm{b}_2}{(2\pi)^4}\,e^{i\bm{k}_T\cdot(\bm{b}_1 - \bm{b}_2)}\expval{\mathcal{O}_{\mathrm{WW}}(\bm{b}_1, \bm{b}_2)}_{Y}\,.
    \label{eq:sx-WW-matching}
\end{equation}
Cyclicity of the color trace gives $\mathcal{O}_{\mathrm{WW}}(\bm{b}_1, \bm{b}_2) = \mathcal{O}_{\mathrm{WW}}(\bm{b}_2, \bm{b}_1)$, establishing the endpoint-exchange symmetry used in the preceding subsection.

For the unpolarized sea-quark distribution, we use the real, $C$-even part of the fundamental dipole correlator~\cite{Hagiwara:2024wqz, Yao:2018vcg},
\begin{equation}
    S_Y(\bm{b}_1, \bm{b}_2) \coloneqq \frac{1}{N_c}\expval{\Re\Tr\left(U_{\bm{b}_1}U_{\bm{b}_2}^\dagger\right)}_{Y}\,.
    \label{eq:sx-fundamental-dipole}
\end{equation}
The imaginary, $C$-odd component describes odderon exchange~\cite{Hatta:2005as} and is not included below.

For the numerical evaluation, we neglect impact-parameter dependence and use rotationally symmetric correlators.  Writing $\bm{b} = (\bm{b}_1 + \bm{b}_2)/2$ and $\bm{r} = \bm{b}_1 - \bm{b}_2$, the center-coordinate integral is replaced by a common effective transverse area, $\int d^2\bm{b} = S_\perp$.  We leave this area unspecified and quote the weighted distributions per unit area,
\begin{equation}
    F_g(Y, k_T) \coloneqq \frac{xf_1^{g, \mathrm{WW}}(x, \bm{k}_T^2)}{S_\perp}\,, \qquad F_q(Y, k_T) \coloneqq \frac{xf_1^{q, \mathrm{sea}}(x, \bm{k}_T^2)}{S_\perp}\,, \qquad k_T = |\bm{k}_T|\,.
    \label{eq:sx-distributions-per-area}
\end{equation}
Both $F_g$ and $F_q$ are dimensionless.  With $\mathcal{W}_Y(r) \coloneqq \expval{\mathcal{O}_{\mathrm{WW}}(\bm{b} + \bm{r}/2, \bm{b} - \bm{r}/2)}_{Y}$, Eq.~\eqref{eq:sx-WW-matching} becomes
\begin{equation}
    F_g(Y, k_T) = \frac{N_c}{4\pi^3\alpha_s}\int_0^\infty dr\,rJ_0(k_Tr)\mathcal{W}_Y(r)\,,
    \label{eq:sx-WW-per-area}
\end{equation}
where $J_0$ is the Bessel function of the first kind of order zero.  Similarly, we write the homogeneous dipole as $S_Y(r)$ and define its Fourier transform by
\begin{equation}
    \widetilde{S}_Y(\ell_T)
    \coloneqq \int\frac{d^2\bm{r}}{(2\pi)^2}\,e^{-i\bm{\ell}_T\cdot\bm{r}}S_Y(r)
    = \frac{1}{2\pi}\int_0^\infty dr\,rJ_0(\ell_Tr)S_Y(r)\,, \qquad
    \ell_T = |\bm{\ell}_T|\,.
    \label{eq:sx-dipole-Fourier}
\end{equation}

The sea-quark distribution is obtained by evaluating the quantum quark correlator in the same classical gluon field at its lowest nonzero order.  For one massless quark flavor with a future-pointing staple, the known small-$x$ expression is~\cite{McLerran:1998nk, Marquet:2009ca}
\begin{equation}
    F_q(Y, k_T) = \frac{N_c}{4\pi^4}\int d^2\bm{\ell}_T\,\widetilde{S}_Y(\ell_T)\mathcal{K}_q(\bm{k}_T, \bm{\ell}_T)\,,
    \label{eq:sx-sea-quark-matching}
\end{equation}
with
\begin{equation}
    \mathcal{K}_q(\bm{k}_T, \bm{\ell}_T) \coloneqq 1 - \frac{\bm{k}_T\cdot(\bm{k}_T - \bm{\ell}_T)}{\bm{k}_T^2 - (\bm{k}_T - \bm{\ell}_T)^2}\ln\frac{\bm{k}_T^2}{(\bm{k}_T - \bm{\ell}_T)^2}\,.
    \label{eq:sx-sea-quark-kernel}
\end{equation}
The longitudinal splitting fraction has already been integrated at leading logarithmic accuracy, while the full transverse-momentum dependence of the kernel is retained.  The angular integral can be performed analytically for the rotationally symmetric input; the resulting one-dimensional representation.

Here, $q$ denotes a single quark flavor and does not include antiquarks.  Within the $C$-even matching used here, the antiquark distribution equals the quark distribution, so the sum over $N_f$ massless flavors of quarks and antiquarks is $2N_fF_q$.  This factor is not included in the single-flavor distributions shown below.  The imaginary, $C$-odd dipole is not included in this input.  Accordingly, the retained sea-quark contribution contains no Sivers term, without implying that the Sivers function of a general polarized sea vanishes.

For the WW gluon and $C$-even sea-quark contributions considered here, the longitudinal and transverse canonical momentum densities are
\begin{equation}
    \frac{\mathcal{T}_a^{++}}{S_\perp} = P^+F_a(Y, k_T)\,, \qquad \frac{\mathcal{T}_a^{+i}}{S_\perp} = \frac{k_T^i}{x}F_a(Y, k_T)\,, \qquad a = g, q\,.
    \label{eq:sx-momentum-densities-per-area}
\end{equation}
For the numerical presentation, we divide the longitudinal component by the target momentum $P^+$, so that $F_a$ represents the longitudinal momentum-fraction density per unit transverse area.

The sea-quark matching in Eq.~\eqref{eq:sx-sea-quark-matching} determines the $\gamma^+$ projection, but not the $\gamma^i$ projection required for $\mathcal{T}_q^{i+}$.  The latter is considered separately through the high-$k_T$ matching in Sec.~\ref{subsec:smallx-quark-tail}.

\subsection{Numerical results for gluon and sea-quark momentum densities}
\label{subsec:smallx-MV-BK}

We now calculate the gluon and sea-quark momentum densities using a common target input specified by a McLerran--Venugopalan (MV) initial condition and fixed-coupling leading-order (LO) Balitsky--Kovchegov (BK) evolution.  Only the fundamental dipole $S_Y(r)$ is evolved numerically.  The sea-quark distribution follows from the matching in the preceding subsection, whereas the WW gluon distribution requires an additional approximation for the Wilson-line correlations.

Writing $S_Y(r) = \exp[-\Gamma_Y(r)]$, we adopt an infrared-regulated McLerran--Venugopalan (MV) initial condition~\cite{McLerran:1993ka}, with the logarithmic prescription used by Ref.~\cite{Caucal:2025xxh},
\begin{equation}
    S_0(r) = \exp[-\Gamma_0(r)]\,, \qquad \Gamma_0(r) =
    \frac{Q_0^2r^2}{4}\ln\left(e + \frac{1}{r\Lambda_{\mathrm{IR}}}\right)\,.
    \label{eq:sx-MV-initial-condition}
\end{equation}
Here, $Q_0^2$ controls the normalization of the initial dipole exponent, while $\Lambda_{\mathrm{IR}}$ sets the infrared scale in the logarithm.  The additive constant $e$ regularizes the logarithm at large $r$.  At small $r$,
\begin{equation}
    1 - S_0(r) \simeq \frac{Q_0^2r^2}{4}\ln\frac{1}{r\Lambda_{\mathrm{IR}}}\,.
\end{equation}
The input parameter $Q_0$ is distinguished from the saturation momentum $Q_s(Y)$ defined below.

Neglecting impact-parameter dependence, the dipole satisfies the fixed-coupling LO BK equation~\cite{Balitsky:1995ub, Kovchegov:1999yj}
\begin{equation}
    \frac{\partial S_Y(r)}{\partial Y} = \frac{\alpha_sN_c}{2\pi^2}\int d^2\bm{r}_1\,\frac{r^2}{r_1^2r_2^2}\left[S_Y(r_1)S_Y(r_2) - S_Y(r)\right]\,, \qquad \bm{r}_2 = \bm{r} - \bm{r}_1\,,
    \label{eq:sx-BK-equation}
\end{equation}
where $r_i = |\bm{r}_i|$.  We keep $\alpha_s$ fixed and use the same coupling in the evolution and the TMD matching prefactors.

The evolved dipole does not by itself determine the WW correlator $\mathcal{W}_Y(r)$.  At each rapidity, we reconstruct this correlator using the nonlinear Gaussian approximation~\cite{Lappi:2017skr, Mantysaari:2019hkq}.  For a homogeneous and rotationally symmetric target, this gives, in the normalization of Eq.~\eqref{eq:sx-WW-operator}~\cite{Dominguez:2011wm},
\begin{equation}
    \mathcal{W}_Y(r) \simeq \frac{1 - e^{- a\Gamma_Y(r)}}{a\Gamma_Y(r)}\nabla_r^2\Gamma_Y(r)\,, \qquad a \coloneqq \frac{C_A}{C_F}\,, \qquad \nabla_r^2 \coloneqq \frac{d^2}{dr^2} + \frac{1}{r}\frac{d}{dr}\,,
    \label{eq:sx-WW-Gaussian-correlator}
\end{equation}
with $C_A = N_c$ and $C_F = (N_c^2 - 1)/(2N_c)$.

Substituting Eq.~\eqref{eq:sx-WW-Gaussian-correlator} into Eq.~\eqref{eq:sx-WW-per-area}, we obtain
\begin{equation}
    F_g(Y, k_T) = \frac{C_F}{4\pi^3\alpha_s}\int_0^\infty dr\,rJ_0(k_Tr)\frac{1 - e^{-a\Gamma_Y(r)}}{\Gamma_Y(r)}\nabla_r^2\Gamma_Y(r)\,.
    \label{eq:sx-WW-Gaussian-distribution}
\end{equation}
All derivatives are taken at fixed $Y$ and retain the full radial dependence of the evolved dipole exponent.  The sea-quark distribution is obtained from the same $S_Y(r)$ through Eqs.~\eqref{eq:sx-dipole-Fourier} and \eqref{eq:sx-sea-quark-matching}.  Both distributions therefore inherit their rapidity dependence from a common dipole, but through different functional relations.

For the numerical illustration, we adopt the proton MV parameter set used by Ref.~\cite{Caucal:2025xxh}, i.e., $Q_0^2 = 0.20~\mathrm{GeV}^2$ and $\Lambda_{\mathrm{IR}} = 0.24~\mathrm{GeV}$.  We assign this dipole to the initial momentum fraction $x_0=10^{-2}$ and evolve it toward smaller $x$ using the fixed-coupling ($\alpha_s = 0.20$) LO BK equation.  We evaluate the distributions at $x = 10^{-2}$, $10^{-3}$, and $10^{-4}$.

The saturation momentum is defined by
\begin{equation}
    S_Y\left(\frac{2}{Q_s(Y)}\right) = e^{-1/2}\,, \qquad \Gamma_Y\left(\frac{2}{Q_s(Y)}\right) = \frac{1}{2}\,.
    \label{eq:sx-saturation-scale}
\end{equation}
Using this condition for both the adopted MV initial condition and the dipoles evolved with the fixed-coupling BK equation, we obtain
\begin{equation}
    Q_s \simeq 0.765,\ 1.077,\ 1.800~\mathrm{GeV} \qquad \text{at} \qquad x = 10^{-2},\ 10^{-3},\ 10^{-4}\,,
\end{equation}
respectively.  The first value corresponds to the initial condition at $x = x_0$.

\begin{figure}
\centering
\begin{minipage}{0.49\columnwidth}
    \centering
    \includegraphics[width=0.99\columnwidth]{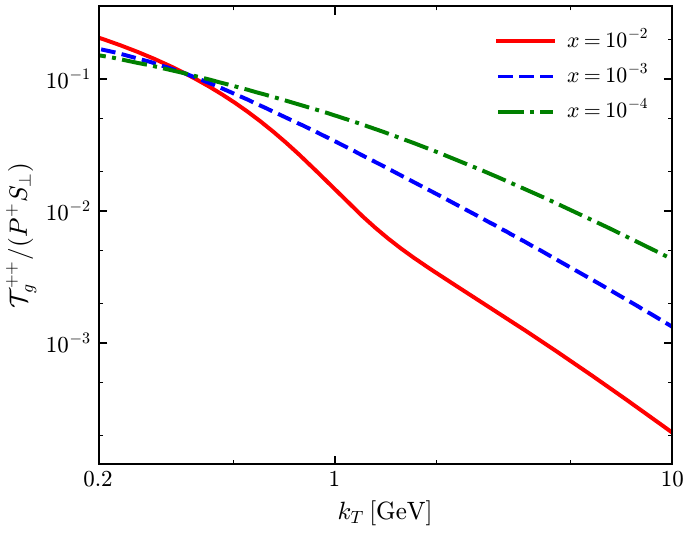}
\end{minipage}
\begin{minipage}{0.49\columnwidth}
    \centering
    \includegraphics[width=0.99\columnwidth]{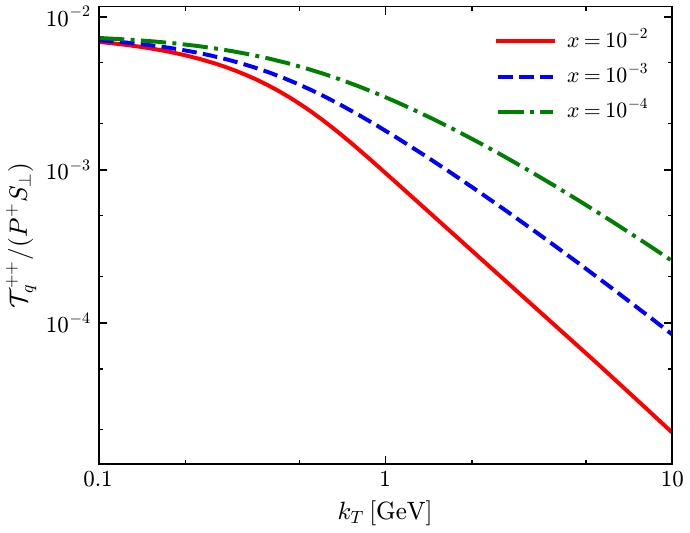}
\end{minipage}
\caption{Longitudinal momentum-fraction densities per unit transverse area, $\mathcal{T}_a^{++}/(P^+S_\perp) = F_a(Y, k_T)$, for gluons (left) and sea quarks (right).  The red solid, blue dashed, and green dash-dotted curves correspond to $x = 10^{-2}$, $10^{-3}$, and $10^{-4}$, respectively.  The curves of gluons are restricted to $k_T \gtrsim 0.2~\mathrm{GeV}$ to exclude the low-momentum region sensitive to the large-$r$ integration limit of the Fourier--Bessel transform, whereas the sea-quark curves extend down to $k_T = 0.1~\mathrm{GeV}$.}
\label{fig:Tpp}
\end{figure}

Figure~\ref{fig:Tpp} shows the longitudinal momentum-fraction densities per unit transverse area, $\mathcal{T}_a^{++}/(P^+S_\perp) = F_a$, for gluons (left panel) and sea quarks (right panel).  Both distributions decrease at large transverse momentum, while their high-$k_T$ regions are enhanced as $x$ decreases.

The decrease of the gluon distribution at large $k_T$ can be understood from the transverse structure of the color field generated by the fast sources.  At transverse momenta well above the saturation scale, the WW correlator admits a dilute approximation.  The dilute-limit behavior can be understood by expanding the Wilson lines in Eq.~\eqref{eq:sx-WW-operator}.  With the line orientation adopted here, $U_{\bm{b}} \simeq 1 - ig\varphi(\bm{b})$, where $\varphi(\bm{b}) = \int dz^-\,\alpha(z^-,\bm{b})$.  Each factor $U^\dagger\partial_iU$ then reduces to $-ig\partial_i\varphi$.  The potential satisfies the transverse Poisson equation $-\nabla_T^2\varphi_a = \bar{\rho}_a$, where $\bar{\rho}_a$ is the longitudinally integrated color source in covariant gauge~\cite{Iancu:2000hn}.  In Fourier space, this gives $\partial_i\varphi_a \to ik_T^i\bar{\rho}_a / \bm{k}_T^2$.  The $1/\bm{k}_T^2$ denominator is the Coulomb Green function in transverse momentum space, while the numerator comes from the transverse derivative.  The transverse factor $k_T^i/\bm{k}_T^2$ scales as $1/k_T$, and its squared magnitude, summed over the transverse components, supplies a $1/k_T^2$ factor in the WW distribution.  This explains the falloff at large $k_T$, while the detailed momentum dependence also reflects the color-source correlations.

The large-$k_T$ decrease of the sea-quark distribution in the right panel of Fig.~\ref{fig:Tpp} reflects the transverse-momentum dependence of $g\to q\bar q$ splitting~\cite{Marquet:2009ca}.  Even when the transverse momentum $\bm{\ell}_T$ entering from the target is small, the quark and antiquark can carry nearly opposite, large transverse momenta.  Such configurations are kinematically allowed, but their weight decreases as the relative transverse momentum increases.  This suppression is explicit after averaging the kernel in Eq.~\eqref{eq:sx-sea-quark-kernel} over the relative azimuthal angle $\phi$ between $\bm{k}_T$ and $\bm{\ell}_T$.  Choosing $\bm{k}_T$ along the first transverse axis, we define the angular average and obtain its large-$k_T$ expansion as
\begin{equation}
\begin{split}
    \overline{\mathcal{K}}_q(\ell_T/k_T) &\coloneqq \frac{1}{2\pi}\int_0^{2\pi}d\phi\,\mathcal{K}_q\left(k_T(1, 0), \ell_T(\cos\phi,\sin\phi)\right) = \frac{\ell_T^2}{3k_T^2} + O\left(\frac{\ell_T^4}{k_T^4}\right)\,, \qquad \ell_T \ll k_T\,.
\end{split}
\label{eq:sx-sea-quark-kernel-large-k}
\end{equation}
The angular average depends only on the ratio $\ell_T / k_T$.  At fixed $\ell_T$, the leading term decreases as $1 / k_T^2$.  Integration over the ordered region therefore produces a decreasing contribution weighted by the corresponding $\ell_T^2$ moment of the dipole Fourier distribution. 

In the low-$k_T$ region, the kernel, $\mathcal{K}_q$, approaches unity as $k_T/\ell_T \to 0$ at fixed nonzero $\ell_T$.  This follows directly from Eq.~\eqref{eq:sx-sea-quark-kernel}.  The normalization of the dipole implies
\begin{equation}
    \int d^2\bm{\ell}_T\,\widetilde{S}_Y(\ell_T) = S_Y(0) = 1\,.
\end{equation}
For the dipole input used here, we therefore obtain~\cite{Marquet:2009ca}
\begin{equation}
    \lim_{k_T\to0}F_q(Y, k_T) = \frac{N_c}{4\pi^4}\,.
    \label{eq:sx-sea-quark-saturation-limit}
\end{equation}
The curves consequently approach a common low-momentum limit.

As $x$ decreases, the saturation momentum $Q_s(Y)$ increases, and the momentum distributions extend toward larger $k_T$~\cite{Iancu:2002tr}.  This provides a qualitative explanation of the enhancement at fixed high $k_T$ observed in Fig.~\ref{fig:Tpp}.

\begin{figure}
\centering
\begin{minipage}{0.49\columnwidth}
    \centering
    \includegraphics[width=0.99\columnwidth]{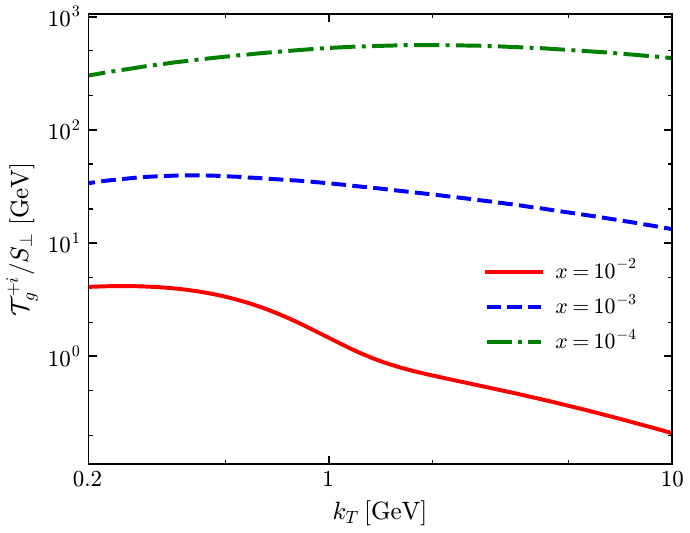}
\end{minipage}
\begin{minipage}{0.49\columnwidth}
    \centering
    \includegraphics[width=0.99\columnwidth]{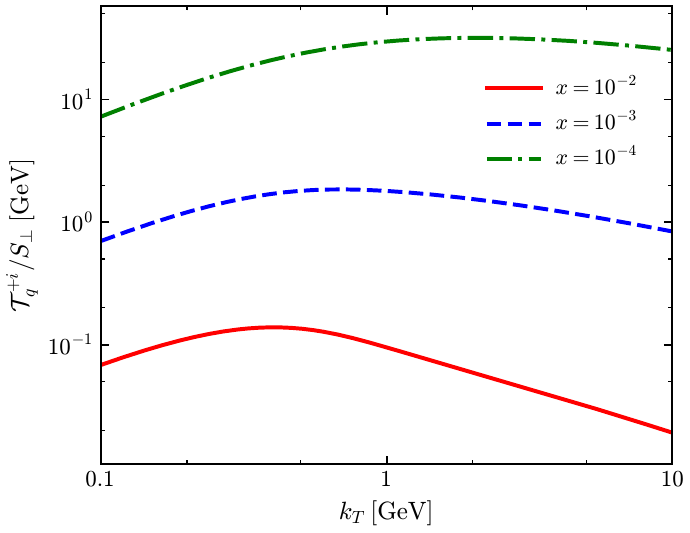}
\end{minipage}
\caption{Magnitudes of the transverse canonical momentum densities per unit transverse area for gluons (left) and sea quarks (right).  Since $\mathcal{T}_a^{+i}$ is directed along $\bm{k}_T$, the plotted quantity is $k_TF_a(Y,k_T)/x$, with $a = g, q$.  The red solid, blue dashed, and green dash-dotted curves correspond to $x = 10^{-2}$, $10^{-3}$, and $10^{-4}$, respectively.  The sea-quark distribution refers to a single massless flavor and does not include antiquarks.  The gluon curves are restricted to $k_T \gtrsim 0.2~\mathrm{GeV}$ to exclude the low-momentum region sensitive to the large-$r$ integration limit of the Fourier--Bessel transform, whereas the sea-quark curves extend down to $k_T = 0.1~\mathrm{GeV}$.
}
\label{fig:Tpi}
\end{figure}

Figure~\ref{fig:Tpi} shows the magnitude of the transverse canonical momentum density per unit transverse area for gluons (left panel) and sea quarks (right panel).  Since $\mathcal{T}_a^{+i}$ is directed along $\bm{k}_T$, the plotted quantity is $k_TF_a(Y, k_T)/x$, with $a = g, q$.  A prominent feature is the large separation between the curves at different $x$, compared with the longitudinal momentum-fraction densities in Fig.~\ref{fig:Tpp}.  This difference follows from the momentum weights entering the two EMT components: the longitudinal momentum density weights each parton by $xP^+$, whereas the transverse momentum density weights it by $k_T^i$.  Where $F_a$ depends only weakly on $Y$, the transverse momentum density consequently grows approximately as $1/x$.  Multiplication by $x$ removes this factor and brings the curves closer together, particularly in the low-$k_T$ sea-quark region where $F_q$ approaches a common saturation limit.  This weighting also has a direct interpretation: since $|dx/dY| = x$, the quantity $x\mathcal{T}_a^{+i}$ is the transverse momentum density per unit logarithmic interval in $x$.  The large separation of the curves should therefore be distinguished from the evolution of the target correlations contained in $F_a$.

\subsection{Quark flux of inertia at high transverse momentum}
\label{subsec:smallx-quark-tail}

For gluons, the eikonal approximation specified above gives $R_v^g = 0$: a nonzero canonical transverse momentum density is not accompanied by a transverse flux of inertia.  This result does not imply the same behavior for sea quarks.  Their distributions are obtained from the quark correlator in the gluon field, and the transverse flux of inertia depends on the twist-three TMD $f^{\perp q}$.  A separate evaluation is therefore needed to establish whether the sea quarks carry a nonzero transverse flux and whether this flux coincides with their canonical transverse momentum density.

A calculation over the full transverse-momentum range would require the corresponding twist-three quark correlator with multiple scattering retained in the dense regime.  Here, we focus on $k_T \gg Q_s(Y), \Lambda_{\mathrm{QCD}}$ and evaluate the $g\to q$ contribution to the perturbative tail of $f^{\perp q}$~\cite{Bacchetta:2008xw, Bacchetta:2019qkv}.  We also give the corresponding tail of $f_1^q$ to establish an analytic comparison between the transverse flux of inertia and the canonical transverse momentum density at the same matching accuracy.  Our aim is to exhibit a flux--momentum ratio different from both $R_v^g = 0$ and the free-particle reference value $R_v^q = 1$ within the approximations specified below.  The numerical results are restricted to the transverse flux of inertia obtained from the twist-three tail; the $f_1^q$ tail is used only for the analytic comparison.

For the unpolarized target considered here, $f_1^q$ and $f^{\perp q}$ denote the sea-quark contributions for a single massless flavor.  Combining Eq.~\eqref{eq:sx-spin-averaged-quark-projections} with the canonical relation $\mathcal{T}_q^{\mu\nu} = k^\nu\Phi_q^{[\gamma^\mu]}$ for $\nu = +, 1, 2$ gives~\cite{Lorce:2023zzg}
\begin{equation}
    \mathcal{T}_q^{i+} = xk_T^if^{\perp q}\,, \qquad
    \mathcal{T}_q^{i+} - \mathcal{T}_q^{+i} = k_T^i\left(xf^{\perp q} - f_1^q\right)\,.
    \label{eq:sx-quark-unpolarized-inertia-flux}
\end{equation}
The ratio between the transverse flux of inertia and the canonical transverse momentum density is characterized by
\begin{equation}
    R_v^q \coloneqq \frac{xf^{\perp q}}{f_1^q}\,, \qquad
    v_{q, T}^i \coloneqq \frac{\mathcal{T}_q^{i+}}{\mathcal{T}_q^{++}} = \frac{k_T^i}{xP^+}R_v^q\,,
    \label{eq:sx-quark-inertia-flux-ratio}
\end{equation}
where $x > 0$.  At nonzero $\bm{k}_T$, $R_v^q = 0$ corresponds to a vanishing transverse flux of inertia, whereas $R_v^q = 1$ gives equality of the two mixed components in the spin-averaged sector.  Neither condition follows from averaging over the target spin.

At the first nonzero order in $\alpha_s$, the $g\to q$ contributions to the perturbative tails are~\cite{Bacchetta:2008xw, Bacchetta:2019qkv}
\begin{equation}
\begin{split}
    \left[f_1^q(x, \bm{k}_T^2)\right]_{g\to q}^{\mathrm{tail}} &= \frac{\alpha_sT_R}{2\pi^2k_T^2}\int_x^1\frac{dz}{z}\,\left[z^2 + (1 - z)^2\right]g_{\mathrm{coll}}(x/z, \mu^2)\,, \\
    \left[xf^{\perp q}(x, \bm{k}_T^2)\right]_{g\to q}^{\mathrm{tail}} &= \frac{\alpha_sT_R}{2\pi^2k_T^2}\int_x^1\frac{dz}{z}\,z(2z - 1)g_{\mathrm{coll}}(x/z, \mu^2)\,,
\end{split}
\label{eq:sx-quark-gluon-channel-tails}
\end{equation}
where $T_R = 1/2$, $g_{\mathrm{coll}}(y, \mu^2)$ is the unpolarized collinear gluon distribution, and $z = x / y$ is the fraction of the parent gluon momentum carried by the quark.  The superscript ``tail'' denotes the retained leading high-$k_T$ contribution, and the scale arguments of the TMDs are suppressed.  Terms matched onto collinear quark distributions are not included.  For the analytic comparison, we construct the ratio below from the two tail expressions in Eq.~\eqref{eq:sx-quark-gluon-channel-tails}, so that its numerator and denominator refer to the same channel and matching accuracy.

It is convenient to introduce the weighted collinear gluon distribution per unit transverse area,
\begin{equation}
    G_{\mathrm{coll}}(y, \mu^2) \coloneqq \frac{yg_{\mathrm{coll}}(y, \mu^2)}{S_\perp}\,.
    \label{eq:sx-weighted-collinear-gluon}
\end{equation}
The ratio for the retained channel then reads
\begin{equation}
    R_{v, g\to q}^{q, \mathrm{tail}} \coloneqq \frac{x\left[f^{\perp q}\right]_{g\to q}^{\mathrm{tail}}}{\left[f_1^q\right]_{g\to q}^{\mathrm{tail}}} = \frac{\displaystyle\int_x^1dz\,z(2z - 1)G_{\mathrm{coll}}(x/z, \mu^2)}{\displaystyle\int_x^1dz\,\left[z^2 + (1 - z)^2\right]G_{\mathrm{coll}}(x/z, \mu^2)}\,.
    \label{eq:sx-quark-tail-ratio-convolution}
\end{equation}
Before further approximation, the ratio depends on the longitudinal shape of the gluon distribution because the two coefficient functions weight the parent momentum fraction differently.

To connect the tail calculation to the common target input used above, we construct an auxiliary dipole-type gluon distribution and its transverse-cutoff integral~\cite{Dominguez:2011wm},
\begin{equation}
\begin{split}
    F_g^{\mathrm{DP}}(Y, \ell_T) &\coloneqq \frac{N_c}{2\pi^2\alpha_s}\ell_T^2\widetilde{S}_Y(\ell_T)\,, \\
    G_Y^{\mathrm{cut}}(\mu^2) &\coloneqq \int_{\bm{\ell}_T^2<\mu^2}d^2\bm{\ell}_T\,F_g^{\mathrm{DP}}(Y, \ell_T) = \frac{N_c}{\pi\alpha_s}\int_0^\mu d\ell_T\,\ell_T^3\widetilde{S}_Y(\ell_T)\,.
\end{split}
\label{eq:sx-cutoff-gluon-input}
\end{equation}
Here, $F_g^{\mathrm{DP}}$ is distinct from the WW distribution entering the gluon EMT.  The quantity $G_Y^{\mathrm{cut}}$ has mass dimension two and is used as a model for $G_{\mathrm{coll}}$; it is not identified with an $\overline{\mathrm{MS}}$ PDF without a separate scheme matching.  Its rapidity dependence is inherited from the same BK-evolved dipole.

Following the treatment of the longitudinal convolution for the unpolarized sea-quark distribution in Ref.~\cite{Marquet:2009ca}, we neglect the dependence of the weighted gluon input on the parent momentum fraction within the convolution.  We apply the same prescription to the twist-three tail, while retaining the full $z$ dependence of both matching coefficients.  Together with the cutoff gluon model introduced above, this gives
\begin{equation}
    G_{\mathrm{coll}}(x/z, \mu^2) \simeq G_{\mathrm{coll}}(x, \mu^2) \longrightarrow G_Y^{\mathrm{cut}}(\mu^2)\,, \qquad \int_x^1dz \longrightarrow \int_0^1dz\,,
    \label{eq:sx-local-longitudinal-approximation}
\end{equation}
where the first step neglects the rapidity shift from $Y$ to $Y + \ln z$, and the second uses our model for the collinear gluon input.  The lower integration limit is set to zero in the small-$x$ approximation.  We choose $\mu = c_hk_T$ with $c_h = 1$ for the numerical results and use the same input and scale in the two tail expressions when forming the analytic ratio.  The longitudinal approximation is distinct from the transverse hierarchy $k_T \gg Q_s(Y)$; its accuracy is not controlled by increasing $k_T$ alone.

The remaining longitudinal integrals are
\begin{equation}
    T_R\int_0^1dz\,\left[z^2 + (1 - z)^2\right] = \frac{1}{3}\,, \qquad
    T_R\int_0^1dz\,z(2z-1) = \frac{1}{12}\,.
    \label{eq:sx-integrated-quark-tail-coefficients}
\end{equation}
The negative contribution of $z(2z - 1)$ for $z < 1 / 2$ is retained.  Equations~\eqref{eq:sx-quark-gluon-channel-tails} and \eqref{eq:sx-local-longitudinal-approximation} consequently give
\begin{equation}
\begin{split}
    \frac{\left[f_1^q\right]_{g\to q}^{\mathrm{tail}}}{S_\perp} &\simeq \frac{\alpha_s}{6\pi^2xk_T^2}G_Y^{\mathrm{cut}}(c_h^2k_T^2)\,, \qquad \frac{\left[f^{\perp q}\right]_{g\to q}^{\mathrm{tail}}}{S_\perp} \simeq \frac{\alpha_s}{24\pi^2x^2k_T^2}G_Y^{\mathrm{cut}}(c_h^2k_T^2)\,.
\end{split}
\label{eq:sx-quark-local-tails}
\end{equation}
Within this prescription, the common gluon input cancels in the ratio, yielding
\begin{equation}
    \left.R_{v,g\to q}^{q,\mathrm{tail}}\right|_{\mathrm{local}} = \frac{1}{4}\,.
    \label{eq:sx-quark-tail-one-quarter}
\end{equation}
Accordingly, the retained sea-quark contribution carries a nonzero transverse flux of inertia equal to one quarter of its canonical transverse momentum density,
\begin{equation}
    \left[\mathcal{T}_q^{i+}\right]_{g\to q}^{\mathrm{tail}} = \frac{1}{4}\left[\mathcal{T}_q^{+i}\right]_{g\to q}^{\mathrm{tail}}\,, \qquad
    \left.v_{q,T}^i\right|_{\mathrm{tail,local}} = \frac{k_T^i}{4xP^+}\,.
    \label{eq:sx-quark-tail-inertia-flux}
\end{equation}
The velocity here is formed from the same tail contributions to the EMT.  This example differs from both the vanishing gluon inertia flux in the eikonal approximation and the equality of momentum density and inertia flux for free particles.

The value $1/4$ is an algebraic consequence of the coefficient functions and the longitudinal approximation in Eq.~\eqref{eq:sx-local-longitudinal-approximation}, not a universal small-$x$ limit or a prediction of BK evolution alone.  Restoring the parent-fraction dependence or the finite lower integration limit generally changes the ratio, as can contributions beyond the retained channel and matching order.  No value of $R_v^q$ in the saturation region is inferred from this result.

\begin{figure}
\centering
\includegraphics[width=0.49\columnwidth]{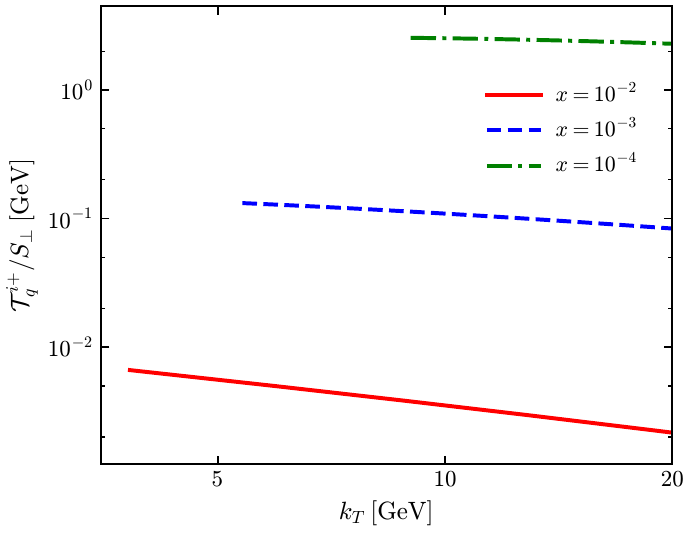}
\caption{Magnitude of the sea-quark transverse flux of inertia per unit transverse area, obtained from the $g\to q$ contribution to the high-$k_T$ tail of $f^{\perp q}$.  Since the flux is directed along $\bm{k}_T$, the plotted quantity is $xk_T[f^{\perp q}]_{g\to q}^{\mathrm{tail}}/S_\perp$.  The red solid, blue dashed, and green dash-dotted curves correspond to $x = 10^{-2}$, $10^{-3}$, and $10^{-4}$, respectively.  The results use the longitudinal approximation in Eq.~\eqref{eq:sx-local-longitudinal-approximation}, the scale $\mu=k_T$, and the cutoff gluon input constructed from the BK-evolved dipole.  Each curve is shown only for $k_T \geq 5Q_s(Y)$ to stay away from the saturation region.  The distribution refers to a single massless quark flavor and does not include antiquarks.
}
\label{fig:Tip}
\end{figure}

We now evaluate the transverse flux of inertia using the twist-three tail and the cutoff gluon input obtained from the common BK-evolved dipole.  Figure~\ref{fig:Tip} shows the magnitude of this flux per unit transverse area for a single massless quark flavor.  With the scale choice $\mu = k_T$, Eq.~\eqref{eq:sx-quark-local-tails} gives
\begin{equation}
    \frac{\left[\mathcal{T}_q^{i+}\right]_{g\to q}^{\mathrm{tail}}}{S_\perp}
    \simeq \frac{k_T^i}{k_T}\frac{\alpha_s}{24\pi^2xk_T}
    G_Y^{\mathrm{cut}}(k_T^2)\,.
    \label{eq:sx-quark-tail-flux-display}
\end{equation}
The plotted quantity is therefore $xk_T[f^{\perp q}]_{g\to q}^{\mathrm{tail}}/S_\perp$.

The sea-quark flux is nonzero and directed along $\bm{k}_T$, in contrast to the vanishing gluon flux in the eikonal approximation considered above.  At fixed $k_T$ within the common displayed range, its magnitude increases as $x$ decreases.  Equation~\eqref{eq:sx-quark-tail-flux-display} separates two sources of this enhancement: the explicit factor $1/x$ and the rapidity dependence of the gluon input $G_Y^{\mathrm{cut}}$.  The separation between the curves therefore reflects both the longitudinal momentum weighting and the evolution of the target correlations.

The flux decreases with $k_T$ over the displayed range, but the decrease becomes weaker after evolution.  Its momentum dependence is governed by the competition between the explicit $1/k_T$ factor and the increase of the cumulative gluon input as its upper integration limit $\mu = k_T$ grows.  The latter partially compensates the former, so the displayed flux need not follow a pure $1/k_T$ behavior.

These results illustrate a nonzero sea-quark flux of inertia within the specified high-$k_T$ approximation.  The analytic relation $R_{v,g\to q}^{q,\mathrm{tail}} = 1/4$ compares two tail expressions evaluated under the same longitudinal approximation; it does not imply that the flux in Fig.~\ref{fig:Tip} equals one quarter of the full-kernel transverse momentum density in Fig.~\ref{fig:Tpi}.

\section{Conclusions}
\label{sec:conclusions}

We have formulated gluon gravitational TMDs using the light-front gic EMT and derived their relations to gluon TMD correlators.  We have expressed the longitudinal and transverse momentum densities, the transverse flux of inertia, and the transverse stresses in terms of the corresponding TMD projections, including the additional scalar contribution to the gluon transverse stress.  Applying this framework at small $x$, we found that the longitudinal and transverse canonical momentum densities were determined by the unpolarized Weizsäcker–Williams gluon distribution, whereas the transverse flux of inertia and transverse stress vanished in the eikonal approximation considered here.  The nonzero canonical transverse momentum density was therefore not accompanied by a transverse flux of inertia at this order.

We have numerically evaluated the unpolarized gluon and sea-quark momentum densities using a common initial condition and fixed-coupling LO BK evolution.  We have examined their transverse-momentum and rapidity dependence and have also evaluated the sea-quark transverse flux of inertia through the $g\to q$ contribution to the high-$k_T$ tail of the twist-three quark TMD.  This contribution yielded a nonzero flux and, under the additional longitudinal approximation adopted here, a flux-to-momentum ratio of $1/4$.  The sea-quark calculation thus provided an example in which the transverse flux of inertia was finite but did not coincide with the canonical transverse momentum density.

The present calculation used a simple implementation of small-$x$ dynamics and can be extended in several directions.  Incorporating running-coupling effects would improve the treatment of the rapidity evolution, while direct JIMWLK evolution of the relevant Wilson-line correlators would allow the WW distribution to be evaluated beyond the Gaussian approximation used here.  A systematic treatment of subeikonal corrections would allow the first nonvanishing contributions to the higher-twist gluon correlators entering $\mathcal{T}_g^{i+}$ and $\mathcal{T}_g^{ij}$ to be investigated, providing access to the transverse flux of inertia, pressure, and shear that vanished in the present approximation.  Extending the small-$x$ calculation to spin-dependent TMDs would further allow the polarization dependence of these mechanical properties to be explored.  For sea quarks, evaluating the twist-three correlator with multiple scattering retained would extend the analysis of the transverse flux of inertia into the saturation region.

Recent developments in small-$x$ evolution~\cite{Caucal:2025mth} and in the calculation of higher-twist~\cite{Altinoluk:2024zom, Mukherjee:2026cte} and spin-dependent~\cite{Kovchegov:2025gcg} TMDs provide a basis for pursuing these extensions.  We expect that applying these developments to the gluon gravitational TMDs formulated in this work will lead to a more complete understanding of mechanical properties in small-$x$ gluon systems.

\begin{acknowledgments}
The author thanks C\'edric~Lorc\'e for useful discussions.  The present work was partially supported by FoPM, WINGS Program, the University of Tokyo and by Japan Society for the Promotion of Science (JSPS) KAKENHI Grant No. 25K24464.
\end{acknowledgments}

\appendix


\bibliography{reference}

\end{document}